\input aa.cmm
  
%
%
 

\def\etal{{\it et al. }}
\def\1{_{\vert}}

\def\ueber#1#2{{\setbox0=\hbox{$#1$}%
  \setbox1=\hbox to\wd0{\hss$\scriptscriptstyle #2$\hss}%
  \offinterlineskip
  \vbox{\box1\kern0.4mm\box0}}{}}

\font\tenib=cmmib10
\font\sevenib=cmmib10 at 7pt
\font\fiveib=cmmib10 at 5pt
\newfam\mitbfam
\textfont\mitbfam=\tenib
\scriptfont\mitbfam=\sevenib
\scriptscriptfont\mitbfam=\fiveib

\mathchardef\bfo="0\the\mitbfam21
\mathchardef\om"0\the\bffam0A
 3


\MAINTITLE{Temporal optimization of Lagrangian perturbation schemes}

\AUTHOR{Georgios Karakatsanis@1@4, Thomas Buchert@2 and Adrian L. Melott@3}

\INSTITUTE{
@1 Max--Planck--Institut f\"ur Astrophysik,
Postfach 1523, D--85740 Garching, Germany
@2 Theoretische Physik, Ludwig--Maximilians--Universit\"at, 
Theresienstr. 37, D--80333 M\"unchen, Germany
@3 Department of Physics and Astronomy, University of Kansas, 
Lawrence, KS 66045, U.S.A.
@4 National Observatory of Athens, 
Lofos Nimfon, Thesio, 18115 Athens, Greece}

\DATE{Received ????, accepted ????} 

\overfullrule=0pt
 
\ABSTRACT{The Lagrangian perturbation theory on Friedmann--Lema\^\i tre
cosmologies is compared with numerical simulations (tree--, adaptive
P$^3$M-- and PM codes).
In previous work we have probed the large--scale performance of the
Lagrangian perturbation solutions up to the third order by
studying their cross--correlations with N--body simulations for various
power spectra (Buchert \etal 1994, Melott \etal 1995, Wei{\ss} \etal
1996). Thereby, spatial optimization techniques were applied by 
(high--frequency--)filtering of the initial power spectra. 
In this work the novel method of temporal optimization 
[Shifted--Time--Approximation (STA) and Frozen--Time--Approximation (FTA)] is
investigated and used. The method is designed to compensate 
the native property of 
Lagrangian perturbation solutions to delay the collapse 
of structures. The method can be treated analytically. 
Applying the STA and FTA prescriptions a significant
improvement of the performance of Lagrangian perturbation schemes up to r.m.s density contrast of about 10 (as 
measured by cross--correlation, relative phase error and power--spectrum statistics) is observed. 
Using this tool we investigate a local study of special
clustering models of dark matter as candidates
for typical elements of the large--scale structure in the Universe,
and so also focus on the performance of the perturbation solutions
on smaller scales at high--spatial resolution.
The models analyzed were presented in (Buchert \etal 1996) and allow studying 
typical features of the clustering process in the non--linear
regime. 
The spatial and temporal limits of applicability of the solutions
at second and third order
are determined and compared with the first--order solution,
which is equivalent to the ``Zel'dovich approximation'' 
(Zel'dovich 1970, 1973) for the type of initial data analyzed.}

 
\THESAURUS{
02.07.1; 02.09.1; 03.13.1; 12.03.4; 12.12.1}
\OFFPRINTS{G. Karakatsanis}

\input mssymb
\input psfig

\maketitle
 
\titlea{Introduction}
 
It is generally appreciated that Lagrangian perturbation theory
provides successful models of large--scale structure down to
the scale where the density field becomes non--linear
(the r.m.s. density contrast is of order unity) (Kofman \etal 1992,
Coles \etal 1993, Melott \etal 1994, Buchert \etal 1994, Bouchet \etal 1995).
For models with considerable small--scale power the truncation of
high--frequency components in the initial fluctuation spectrum
allows application of the Lagrangian schemes down to
galaxy group mass scales as was found for a family of power--law
hierarchical models (Melott \etal 1994, 1995).
The Lagrangian schemes are most suitable tools in the regime where
the spectral index is negative on small scales. In the case
where the truncation scale in the initial spectrum corresponds to the
Nyquist frequency of a N--body simulation there is
no need for N--body computing. 
There is agreement about the fact that Lagrangian schemes can replace
N--body integrators above some scale close to, but smaller than 
the non--linearity scale; they provide
fast and effective one--time step mappings applicable to various
kinds of studies of hierarchical cosmogonies (such as CDM models, Wei{\ss}
\etal 1996) and statistical studies of, e.g., 
the modeling of pencilbeams at high resolution (Wei{\ss} \& Buchert 1993), or 
the distribution of clusters (Borgani \etal 1995).
In the previously mentioned works the decreasing performance of the spatially
optimized Lagrangian schemes at nonlinear scales and, 
of course, near the epoch of shell-crossing has been
pointed out. Here we show the possibility to overcome this problem and
to maintain a good performance until the epoch of caustic
formation by employing a temporal optimization method 
(Shifted--Time--Approximation). Further we suggest to extend 
the usability of the
Lagrangian perturbation theory to even later stages where the
Lagrangian theory is not formally valid using the Frozen--Time--Approximation.

The statement of applicability of the Lagrangian approximations
in previous works is made on the basis of cross--correlation statistics of density fields
in which the internal substructures are not resolved. 
We here also address the question whether these approximations can model
these substructures. Since analytical models are much faster to execute,
it is our goal to understand whether and 
how these substructures compare with those of N--body simulations.
Here, we should be able to learn
more about the details of the clustering process, but also about the problems
which are inherent in a Lagrangian perturbation
approach.  
With this work we want to approach the limits of Lagrangian perturbation 
schemes by means of studying special initial data which are suitable
to process these questions efficiently.

On one hand we have taken various numerical N--body
integrators to assure that the features we want to compare with
do not depend on whether we use a tree--code, an adaptive P$^3$M--code, or
a PM--code.
Previous comparisons have only been performed with PM--codes.
On the other hand we are interested in both the local details of structure
formation and the statistical properties of the overall distribution, 
which have been tested in previous work for (physically) 
larger simulation boxes.

\titlea{Clustering models, N--body integrators and cross--correlation statistics}

\titleb{Clustering models}
We start with the analysis of a simple plane--wave model (Model I)
as described by Buchert \etal (1996).
We stick to that model first since, despite its simplicity, it
already shows the principal features of gravitational
collapse we are interested in. Details about the construction of this 
model are given in the appendix of Buchert \etal (1996).
Also in other work on related subjects this model is useful as
an example (Mo \& Buchert 1990,
Matarrese \etal 1992), or as a toy--model for the comparison of different
approximation schemes. 

We then move to a generic model (Model II), i.e., a model without symmetry,
but restricted to a small enough box to assure the resolution of
patterns we are interested in. The construction and the properties
of this model are also 
described in (Buchert \etal 1996). 
In both the special and the generic
model we quantitatively
investigate the delay of collapse times in the Lagrangian schemes compared with
the collapse time of the numerically simulated structures,
and express this delay in terms of the r.m.s. density fluctuation of structures or
the spatial scale, respectively.
\titleb{N--body integrators}
We use a hierarchical tree--code (Bouchet \& Hernquist 1988) with
incorporated periodic boundary conditions based on the Ewald
method (Hernquist \etal 1991) as well as the adaptive P$^3$M--code
by Couchman (1991), which is also used as a standard PM--code as
described in (Couchman 1991).

The simulations have been done for $64^3$ particles and the standard
choices for the tolerance parameter $0.75$, softening--length $0.015$ and time--step $0.2$
in the tree--code
(Suginohara \& Suto, priv.comm., Suginohara \etal 1991). For the
P$^3$M- and the PM code the settings are according to the work of
Efstathiou et al. (1985). Grid spacing for the PM simulations and initially for P$^3$M was $128^3$.
The parameter settings used have been tested on the 
exact plane--symmetric model to yield the
same collapse time.

\titleb{The statistics used}

We use three different statistical methods.
Firstly, the cross--correlation coefficient $S$ to compare
the resulting density fields,
$$
S := {<(\delta_1 \delta_2)> \over \sigma_1 \sigma_2} \;\;, \eqno(1)
$$
where $\delta_{\ell}, \ell=1,2$  represent the r.m.s. density contrasts in
the analytical and the numerical approximations, respectively,
$\sigma_{\ell} = \sqrt{<\delta_{\ell}^2>-<\delta_{\ell}>^2}$
is the standard deviation in a Gaussian random field;
averages $<...>$ are taken over the entire distribution. 
We will plot $S$ as a function of $\sigma_2$ for different Gaussian 
smoothing lengths. 

\noindent
We believe this
is the most important statistical test, because it measures whether the
approximation is moving mass to the right place, with an emphasis on
dense regions. (We plan to address bulk flows in underdense 
regions in a future study.) 
We allow for small errors by presenting $S$ for the two
density arrays smoothed at a variety of smoothing lengths.
The correlation coefficient obeys the bound $\vert S\vert \le 1$;
$S = 1$ implies that $\delta_1 = {\cal C} \delta_2$, with $\cal C$
constant.

\smallskip

Secondly, the power spectrum of the evolved N--body model and the
analytical approximations were calculated and plotted against the wave number
$k:=|\vec k|$. 

Thirdly, the phase angle accuracy is measured and displayed by
$<\cos\theta>_k$, where $\theta=\phi_1-\phi_2$ is the difference in the
phase angle of the Fourier coefficients of mass density between the
approximation and the simulation. 
We present the results on the relative phase errors in terms of
$\cos(\theta)$ as a function of $k$ (calculated in spherical shells in
$k-$space).
Perfect agreement between the N-body result
and the analytical scheme implies $\cos(\theta)=1$, anti-correlated
phases have $\cos(\theta)=-1$, and for randomized phases $\cos(\theta)$
would average to $0$.

For further details on these statistics we
refer the reader to Melott \etal (1994) and Buchert \etal 
(1994).

\titlea{Results and Discussion}

\titleb{Comparison between different numerical integrators}

In Fig.1a we present a comparison of the density fields at three
evolution stages as predicted by the tree--code, the adaptive
P$^3$M--code and the PM--code for Model I.
Since the coincidence of all three
algorithms is excellent (the cross correlation coefficient lies
between 0.95 and 1.00 and the relative phase error between 0.7 and 1.00), we henceforth stick to the adaptive
P$^3$M--code, which is about 2 times  faster than the tree--code and
has a significantly higher force resolution -- due to the mesh refinement at
dense regions -- than the faster PM-code.
This does, however, not imply that AP$^3$M is more accurate than PM
(see Suisalu \& Saar 1996, Melott \etal 1996).
In Fig.1b some statistics of this comparison are displayed.

\titleb{The temporal optimization methods}

Various works on the comparison of the Lagrangian perturbation theory
with numerical simulations (Melott \etal 1994, Buchert \etal 1994, 
Munshi \etal 1994, Bouchet \etal 1995)
show an improvement of the performance of the Lagrangian perturbation
theory using the second--order scheme independent of the
initial conditions and the fluctuation scale. This fact concerns not
only the spatial accuracy mirrored in the cross--correlations, but also
the time accuracy concerning the collapse time as compared with the
numerical simulations. We are led to the conclusion that the spatial accuracy is also a consequence of the time evolution accuracy.
The collapse time accuracy of the Lagrangian schemes increases
with increasing order until the epoch of shell crossing. The time coefficients of the Lagrangian theory grow
proportional to $a^{n}$ for a given order n. This also means that the
higher the order the earlier the break down of the theory (``blow--up effect'').
During this work we indeed
observed that time--shifted low--order schemes produce configurations
similar to those of unshifted higher--order schemes suggesting that the
Lagrangian theory reproduces the systems evolution correctly but delayed. Under the assumption that the complete perturbation series converges to the temporally (and spatially) correct solution we can thus test 
the hypothesis that time shifting of the available low--order schemes will 
mimick the higher--order effects leading to optimal results 
compared with numerical simulations.
This means that one can compare 
the Lagrangian perturbation theory with numerical simulations 
at the same expansion factor only formally, 
because the evolution stages don't correspond physically.

\smallskip

We introduce the time--shift factors
s$_{n}(a_{num})$ in order to quantify the amount of the time-shift for
a given order n matching the numerical simulation at the expansion factor $a_{num}$.
Formally the assumption reads:

$a_{n}=a_{num} \cdot s_{n}(a_{num})$ with

$s_{n}<s_{n-1}$ ; $s_{n}>1$ ; $s_{\infty}$=1 ; $a_{n}\le a_{n}^{crit}\;\;,$

$a_{n}$ is the corresponding optimally shifted expansion factor for the
order n, and $a_{n}^{crit}$ is the expansion factor of the shell--crossing stage
which can be calculated analytically.

It turns out that this method of the Shifted--Time--Ap\-proxi\-mation (STA) 
leads to an astonishingly good agreement
between the shifted Lagrangian schemes and the numerical simulations
for both models analyzed.

The optimal time--shift has been first determined by minimizing the error
in the cross--correlation statistics. The analysis of the results showed that
the mechanism and criterion of this optimization method is based on the 
r.m.s. density contrast which 
has to be equal (up to about $2 \%$) to that of the numerical 
simulations at the corresponding
stages. This is illustrated in Figure 9 (for Model I as an example), 
where the r.m.s. density 
contrast for the numerical simulation and for both the optimized and 
unoptimized Lagrangian schemes is plotted as a function of the smoothing scale.   
The mechanism of the STA is illustrated in Fig.2a in comparison with 
conventional methods.
The optimal time--shift of the Lagrangian schemes is {\it unique} 
and is determined by the adaptation of the value of the r.m.s. density 
contrast to the numerical value. The error in the cross--correlation 
coefficient, the power--spectrum
and the phase accuracy displays {\it distinct and unique} 
minima which nearly lie at the same
values of $a_{min}$, i.e., the ``optimal'' time--shift predicted is quite close
in all statistics.
Here, it is to be mentioned that the spatial optimization done in the 
``truncated
Zel'dovich--approximation'' (TZA; Melott \etal 1994) is based on the 
cross--correlation coefficient alone. 
However, they found the same rank ordering for phase accuracy as for 
cross--correlation. Further time--shifting 
decreases the performance as shown in the plots of the approximation 
error as a function of the shift amount (Figure 5).

\noindent
Turning this result around we can use the STA method independently of the 
numerical simulations in order to simulate analytically the structure formation process: if a value for the r.m.s.
density contrast on a certain scale for the stage we want to simulate
is given, then we can analytically or even graphically determine if the
STA method is valid (this means if the desired value of the r.m.s. density 
contrast can be reached by the Lagrangian 
perturbation theory), and then use 
the more accurate order and the corresponding optimal expansion factor.

The ``blow--up effect'' signalizes the validity limit of the
Lagrangian perturbation schemes and thus of the STA method: the
structures built decay and the r.m.s. density contrast decreases
rapidly in contrast to the numerical simulations where further 
shell--crossings (due to self--gravitation of multi--stream systems)
hold the structures together.
So we are led to the hypothesis that the last stable configuration 
produced by the Lagrangian
perturbation theory (just before shell--crossing) provides the structural frame for the further nonlinear
evolution. It turned out that this assumption holds.
It means that, in order to get reasonable results even if the 
Lagrangian theory is formally no longer valid, 
one needs to shift the Lagrangian schemes backwards to the analytically 
calculable expansion factor $a_{n}^{crit}$. 
This method we call Frozen--Time--Approximation (FTA).
FTA can also be
treated analytically and leads to very good results for the epochs
shortly after shell--crossing where the Lagrangian perturbation theory is
formally not valid. The mechanism of the FTA method -- similar to the
STA mechanism -- is based on the minimization of the r.m.s. 
density contrast difference between the numerical simulations 
and the Lagrangian schemes. 
(Equality is in this case not possible due to the structure decay 
and the accompanying decrease of the r.m.s. density contrast.) 
This method is explained in Fig.2b.
 
\titleb{Illustration of the optimization results}

Both optimization methods improve significantly on the performance of 
the Lagrangian schemes even at stages where the r.m.s. density contrast is 
above 10. This is shown in Fig.3 for the cross--correlation 
coefficient within the framework of the STA. After the
shell--crossing expansion factor, which signalizes the validity limit for the 
STA, the error increases rapidly.

\hskip -1 true cm
\leftline{\psfig{figure=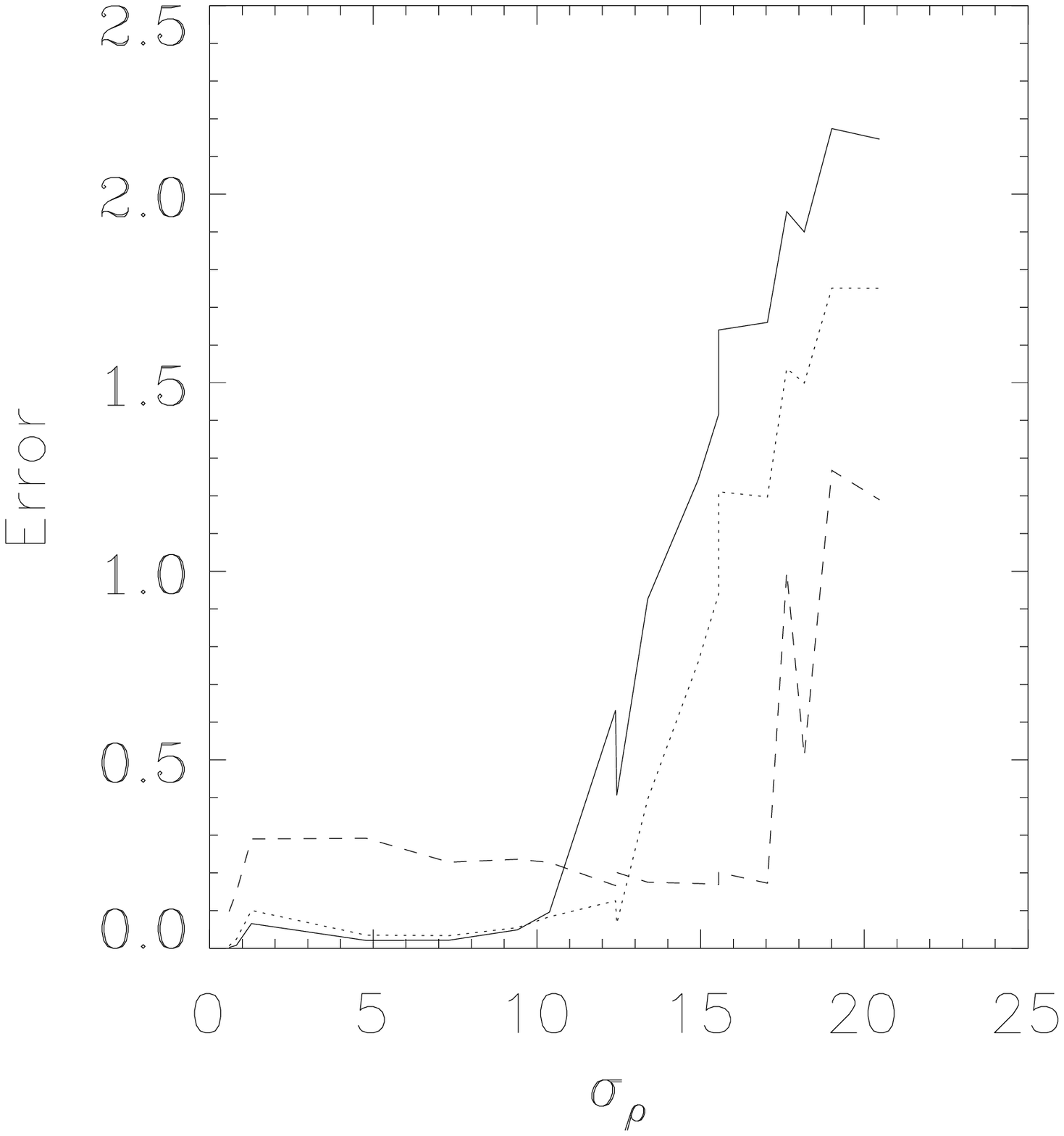,height=5.0cm,width=5.0cm}}
\vskip -5 true cm 
\rightline{\psfig{figure=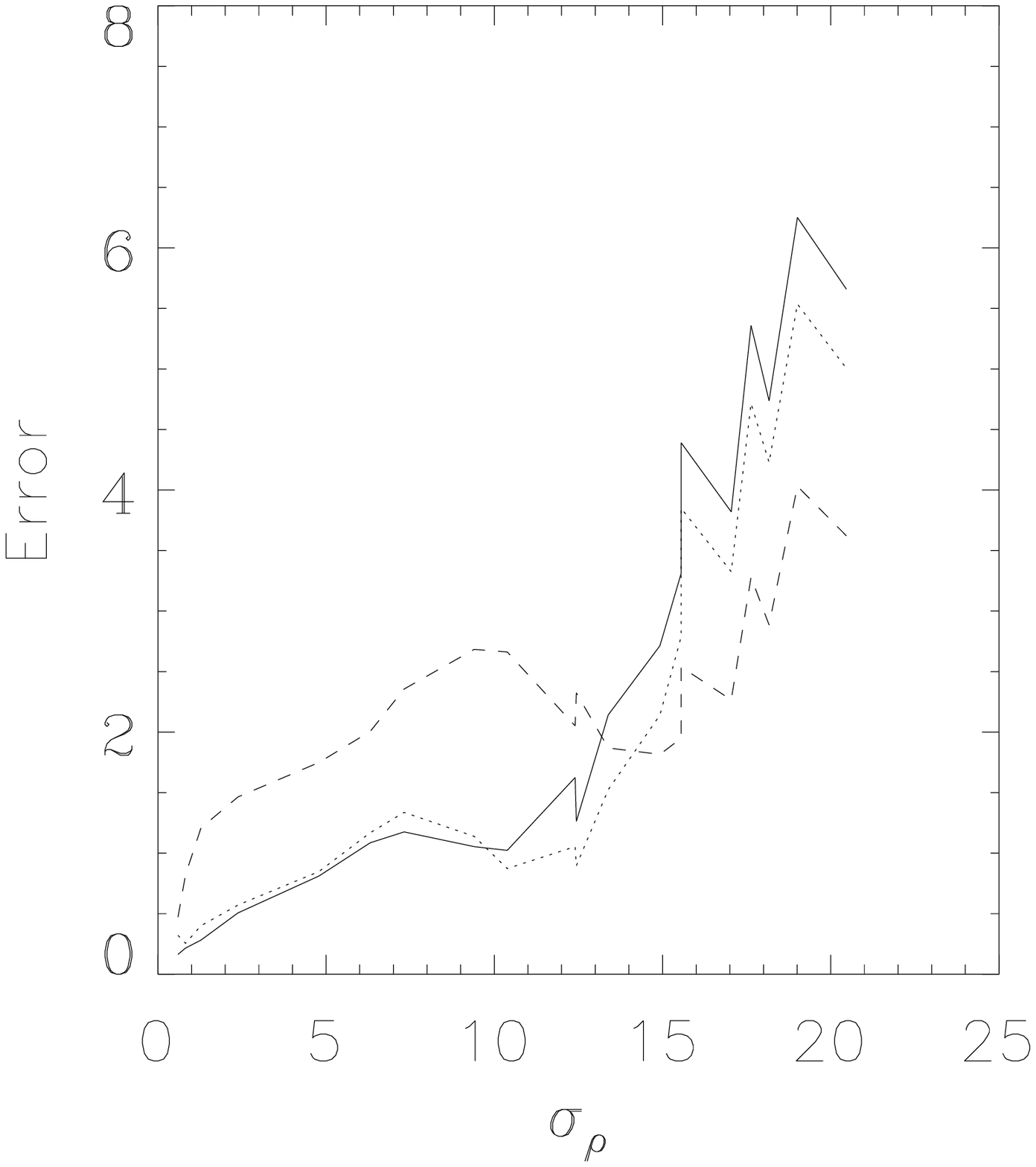,height=5.0cm,width=5.0cm}}
{\bf Figure 3:} The absolute errors in the cross--correlation coefficient
for Model I (left) and Model II (right panel)
between the numerical simulation and the optimized Lagrangian schemes is shown.
The 3rd--order scheme is plotted as a full line, the 2nd--order one as 
a dotted line and 1st--order as a dashed line.

\medskip

The shell--crossing expansion factor for each order
$n$ signalizes the end of the validity of the STA and simultaneously the
onset of the validity regime of the FTA. Thus, both methods can be
combined resulting in an optimal approximation error as
shown in Figs.4a,b.

\medskip

Figure 6 shows the quantitative difference between the optimized and
unoptimized approximation errors in all statistics used over the whole 
evolution sequence for Model II as an example.
Figs.7--10 (some for Model I and some for Model II) further quantify
the gain in performance due to the optimization technique STA 
by depicting the different statistics (described in Subsection
2.3). These statistics show the comparison between the Lagrangian schemes, which 
are evolved to the same expansion factor $a_{num}$
as the N--body run, the N--body result itself, and the 
optimally time--shifted schemes for the STA technique. 

\noindent
The figures together with the captions are self--explanatory, so we
do not present an additional discussion.

\smallskip

Figs. 11a,b show the approximation error from the cross correlation 
and relative phase error statistics
resulting from the application of the FTA method for Models I and II, 
respectively: The approximation error is 4--5 times smaller than 
in the unoptimized case and reaches unique minima at the shell--crossing 
expansion factor. Finally, in Figure 12 we depict slices 
of the density field of 
Model I at an epoch after shell--crossing, 
where the unshifted and shifted Lagrangian schemes are compared 
with the N-body result. It is remarkable that even the 
first--order (Zel'dovich--)approximation displays a large 
gain in performance as demonstrated in Fig.12 by an increase 
of the maximal density 
contrast from 26 to 511 compared to the N--body value of 510.

\titlea{Conclusions}

We presented the novel temporal optimization methods (STA and FTA) based on
native properties of the Lagrangian perturbation schemes which, combined,
allow the analytical simulation of the gravitational evolution
into the nonlinear regime with an optimally minimized approximation
error compared with numerical simulations. We appreciate a significant 
gain in performance, even in the non--linear regime 
(up to r.m.s. density contrasts of about 10) as measured by three 
different statistical methods. 
The advantages of the temporal optimization methods are the 
possibility of a physical comparison between Lagrangian perturbation schemes
and N--body simulations, the use of the full
performance of each Lagrangian order available today, 
the universality of the methods (i.e., their mechanisms are only based
on the variance of the approximated realization), and their 
analytical or even graphical treatment which is easy to put into practice. 

\smallskip
The present results indicate that the {\it physical} content
(concerning the spatial distribution of structure) 
inherent in N--body realizations does not exceed that of Lagrangian 
perturbation schemes up to the abovementioned stages of nonlinearity
and for the given resolution limits. This could be understood in the case
of convergence of the Lagrangian perturbation scheme
to the solution modeled by the N--body simulation, i.e., 1. increasing the
order of the perturbation scheme allows to access smaller spatial scales
(in the limits we have tested), and 2. the main effect of the 4$^{th}$ 
and higher--order corrections before shell--crossing
on the resolved scale can be assigned to  
acceleration of the collapse process which is fully 
compensated for by the applied 
temporal shift. 

\smallskip

As in previous work we also appreciate that the second--order scheme improves
on the first--order scheme (the ``Zel'do\-vich approximation'') 
for both models tested and at all stages within the 
temporal range where the optimization techniques apply; for the 
third--order scheme an improvement has been detected, but this improvement
is negligible at this resolution compared to the gain in performance by going 
to second order. This result might change by going to higher spatial 
resolution. However, we emphasize that the ``blow--up effect'', i.e. the 
validity limit of the n$^{th}$--order perturbation scheme 
imposed by its power law time dependence $\propto a(t)^n$, implies that 
the temporal range of validity
of the schemes will decrease drastically by going to higher orders. 
Therefore, we expect that
going to higher than 3$^{rd}$ order will not be useful. 

\acknow{We would like to thank Fran\c cois Bouchet, Yasushi Suto and Tatsushi
Suginohara for useful comments and discussions 
and for providing us their tree--codes and Hugh Couchman for
providing his adaptive P$^3$M--code.
To Gerhard B\"orner and 
Arno Wei\ss\ we are thankful for
helpful discussions.
\noindent
GK and TB were both supported by 
the ``Sonderforschungsbereich 375 f\"ur
Astro--Teilchenphysik der Deutschen Forschungsgemeinschaft''.
ALM gratefully acknowledges financial support from NASA grant NAGW3832, and
the NSF EPSCoR program, as well as the use of the National Center for
Supercomputing Applications.
}

\begref{References}

\ref
Borgani S., Plionis M., Coles P., Moscardini L., 1995,  MNRAS 277, 1191
\ref
Bouchet F.R., Hernquist L., 1988, ApJ Suppl 68, 521
\ref
Bouchet F.R., Colombi S., Hivon E., Juszkiewicz R., 1995, Astron.
Astrophys. 296, 575
\ref
Buchert T., Melott A.L., Wei\ss\ A.G., 1994, Astron.
Astrophys. 288, 349
\ref
Buchert T., Karakatsanis G., Klaffl R., Schiller P., 1996, Astron.
Astrophys., in press
\ref
Coles P., Melott A.L., Shandarin S.F., 1993, MNRAS 260, 765
\ref
Couchman H.M.P., 1991, Ap.J. 368, L23
\ref
Efstathiou G., Davis M, Frenk C.S., White S.D.M. (1985), ApJ Suppl. 57, 241 
\ref
Hernquist L., Bouchet F.R., Suto Y., 1991, ApJ Suppl. 75, 231
\ref
Kofman L.A., Pogosyan D.Yu., Shandarin S.F., Melott A.L., 1992, ApJ 393, 437 
\ref
Matarrese S., Lucchin F., Moscardini L., Saez D., 1992, MNRAS 259, 437
\ref
Melott A.L., Pellman T., Shandarin S.F., 1994, MNRAS 269, 626 
\ref
Melott A.L., Buchert T., Wei\ss\ A.G., 1995, Astron. Astrophys. 294, 345
\ref
Melott A.L., Splinter R.J., Shandarin S.F., Suto Y., 1996, preprint
\ref
Mo H.J., Buchert T., 1990, Astron. Astrophys. 234, 5
\ref
Moutarde F., Alimi J.-M., Bouchet F.R., Pellat R., Ramani A., 1991, 
ApJ 382, 377
\ref
Munshi D., Sahni V., Starobinsky A.A., 1994, ApJ 436, 517
\ref
Suginohara T., Suto Y., Bouchet F.R., Hernquist L., 1991, ApJ Suppl 75, 631
\ref
Suisalu I., Saar E., 1996, MNRAS, submitted
\ref
Wei{\ss} A.G., Buchert T., 1993, Astron. Astrophys. 274, 1
\ref
Wei{\ss} A.G., Gottl\"ober S., Buchert T., 1996, MNRAS 278, 953
\ref
Zel'dovich Ya.B., 1970, Astron. Astrophys. 5, 84
\ref
Zel'dovich Ya.B., 1973, Astrophysics 6, 164

\endref

\vfill\eject

\noindent
{\bf Figure Captions}
\bigskip\medskip
\noindent
{\bf Figure 1a:} Central slices of the density field 
($64^3$ trajectories collected into a $64^3$ pixel grid) as predicted by 
the PM--Code (top), Tree--Code (middle) and AP$^3$M--Code (bottom panel) 
at expansion factors $a = 3.71$ (left) and $a=5.37$ (right).

\bigskip\noindent
{\bf Figure 1b:} Cross--correlation coefficient and phase angle accuracy 
are depicted for some evolution stages; the
code of reference was AP$^{3}$M. Full lines mark the cross--correlation with
the Tree--Code, dotted lines that with the PM--code
The power--spectrum is also shown: full line: AP$^{3}$M--code, 
dotted line: tree--code, dashed line: PM--code.

\bigskip\noindent
{\bf Figure 2a:} The STA mechanism (Model I). 
The r.m.s. density contrast as a function of the expansion factor is plotted. 
The numerical result is shown as a full line, 3rd order as a dotted line, 
2nd order as a dashed line and 1st order as a dashed--dotted line. 
The best optimization results are obtained when the numerical r.m.s. 
density contrast  is equal to the corresponding value of the Lagrangian 
schemes.

\bigskip\noindent
{\bf Figure 2b:} The FTA mechanism (Model I). The r.m.s. density contrast 
as a function of the expansion factor is plotted. The numerical result is 
shown as a full line, 3rd order as a dotted line, 2nd order as a dashed line 
and 1st order as a dashed--dotted line.

\bigskip\noindent
{\bf Figure 4a:} The combination of STA and FTA (Model I, 2nd order). 
The approximation error as a function of the expansion factor is plotted. 
STA and FTA optimized: full line, without optimization: dotted line.

\bigskip\noindent
{\bf Figure 4b:} The validity regions of STA and FTA (Model II). The r.m.s. 
density contrast as a function of the expansion factor is plotted. 
The numerical result is shown as a full line, 3rd order as a dotted line, 
2nd order as a dashed line and 1st order as a dashed--dotted line.

\bigskip\noindent
{\bf Figure 5:} The absolute error in the magnitudes of the cross--correlation
coefficient (upper), the power--spectrum (middle) and the phases (lower row)
between the numerical result at time $a_{num}$ and Model I
is depicted as a function of $a$; 3rd order is shown as a full line, 2nd order
as a dotted line and 1st order as a dashed line.

\bigskip\noindent
{\bf Figure 6:}
The mean quadratic error in the cross--correlation coefficient
(top row), the power--spectrum (middle row) and the phase--angle (bottom row)
between Model II and the AP$^3$M simulation
is shown as a function of the density contrast of the numerical simulation
for the third--order (first column), second--order 
(second column) and the first--order approximation (third column).
Dotted lines mark the formal comparison at the same expansion factor,
full lines mark the comparison with the ``optimally time--shifted'' 
approximations.

\bigskip\noindent
{\bf Figure 7:} 
The cross--correlation coefficient 
between the numerical result at time $a_{num}$ and Model I at the same 
expansion factor (left panels) and at the optimially shifted time 
(right panels) is depicted as a function of $\sigma_{\rho}$; 3rd order is 
shown as a full line, 2nd order as a dotted line and 1st order as a dashed line.

\bigskip\noindent
{\bf Figure 8:} 
The power--spectrum 
of the numerical result at time $a_{num}$ and Model II at the same expansion 
factor (left panels) and at the optimially shifted time (right panels) 
is depicted as a function of wave number; the numerical result is shown as a 
full line, 3rd order as a dotted line, 2nd order as a dashed line and 
1st order as a dashed--dotted line.

\bigskip\noindent
{\bf Figure 9:} The density contrast of the numerical result at time $a_{num}$ 
and Model I at the same expansion factor (left panels) and at the optimially 
shifted time (right panels) is depicted as a function of smoothing scale 
(in grid units); the numerical result is shown as a full line, 3rd order as a 
dotted line, 2nd order as a dashed line and 1st order as a dashed--dotted line.
It is obvious that the mechanism of STA results in the adaptation of the r.m.s. 
density contrast.

\bigskip\noindent
{\bf Figure 10:} 
The phase--error
between the numerical result at time $a_{num}$ and Model I at the same 
expansion factor (left panels) and at the optimally shifted time (right panels)
is depicted as a function of wavenumber; 3rd order is shown as a full line, 
2nd order as a dotted line and 1st order as a dashed line.
 
\bigskip\noindent
{\bf Figure 11a:} 
The approximation error (Model I) is plotted as a function of the 
expansion factor
for FTA and for different numerical expansion factors: The minimum at the 
expansion factor of shell--crossing for all examined
stages can be seen. In the left panels the approximation error for the
cross--correlation coefficient (increasing order from top to bottom)
and in the right panels the same situation for the relative phase
error are shown. The different lines belong to the different initial
stages in the range from 3.71 to 6.80.

\bigskip\noindent
{\bf Figure 11b:} 
Same as Figure 11a, but for Model II.

\bigskip\noindent
{\bf Figure 12:} Central slices of the density field for
$a_{num} = 3.07$ in Model I ($64^3$ trajectories collected into a $64^3$ pixel 
grid) as predicted by the AP$^3$M--Code (top left) in comparison with the 
STA--optimized 3rd order (top right), the STA--optimized 2nd order (middle left),
the STA--optimized 1st order (middle right) and the unoptimized 2nd and
1st orders (bottom left and right respectively).

\vfill\eject
\bye